\documentclass[copyright,creativecommons]{eptcs}
\providecommand{\event}{LFMTP 2023}

\usepackage{underscore}    
\usepackage[T1]{fontenc}

\usepackage{import} 

\usepackage{amssymb}
\usepackage{amsmath}
\usepackage{amsfonts}

\DeclareMathAlphabet{\mathcal}{OMS}{cmsy}{m}{n}

\usepackage{listings} 
\usepackage[dvipsnames]{xcolor}

\usepackage{xspace}

\usepackage{proof} 

\usepackage{multicol} 
\usepackage{stmaryrd} 

\usepackage[numbers]{natbib}  

\usepackage[nolist]{acronym}

\newtheorem{theorem}[subsubsection]{Theorem}

\lstdefinelanguage{Beluga}{%
  alsodigit = {-},
  morecomment = [l][\color{Gray}]{\%},
  morekeywords = [1]{
    LF, type, ctype, schema, rec, case, of, and,
    some, block, total, fn, let, in, mlam,
    LFR, sort, csort, world,
    partial,
  },
  keywordstyle = [1]{\ttfamily\bfseries\color{violet}},
  morekeywords = [2]{
    lam, app,                               
    e-lam, e-app, e-refl, e-sym, e-trans,   
  },
  keywordstyle = [2]{\ttfamily\color{BrickRed}},
  morekeywords = [3]{
    tp, tm, deq,
  },
  keywordstyle = [3]{\ttfamily\color{NavyBlue}},
  morekeywords = [4]{
    aeq,
  },
  keywordstyle = [4]{\ttfamily\color{Green}},
  morekeywords = [5]{
    xeW,
  },
  keywordstyle = [5]{\ttfamily\color{Orange}},
  morekeywords = [6]{
    xaG,
  },
  keywordstyle = [6]{\ttfamily\color{OliveGreen}},
  morekeywords = [7]{
    xdG, daG,
  },
  keywordstyle = [7]{\ttfamily\color{Blue}},
  alsoletter=/,%
  columns=flexible,%
  sensitive = true,%
  basicstyle=\ttfamily,%
  mathescape=true,%
  escapechar={*},
  literate={%
    {→}{$\rightarrow$ }1%
    {⇒}{$\Rightarrow$ }1%
    {->}{$\rightarrow$ }1%
    {\\}{$\lambda$}1%
    {⊢}{$\vdash\;$}1%
    {|-}{$\vdash\;$}1%
    {sigma}{$\sigma$}1%
  }%
}%

\begin{document}
\title{Contextual Refinement Types}

\author{
  Antoine Gaulin
  \institute{McGill University}
  \email{antoine.gaulin@mail.mcgill.ca}
\and
  Brigitte Pientka
  \institute{McGill University}
  \email{bpientka@cs.mcgill.ca}
}
\def\titlerunning{Contextual Refinement Types}
\def\authorrunning{A. Gaulin, B. Pientka}

\begin{acronym}
  \acro{CMTT}{contextual modal type theory}
  \acro{OL}{object language}
  \acro{HOAS}{higher-order abstract syntax}
\end{acronym}

\maketitle

\begin{abstract}
  We develop an extension of the proof environment \textsc{Beluga} with datasort refinement types and study its impact on mechanized proofs.
  In particular, we introduce \textit{refinement schemas}, which provide fine-grained classification for the structures of contexts and binders.
  Refinement schemas are helpful in concisely representing certain proofs that rely on relations between contexts.
  Our formulation of refinements combines the type checking and sort checking phases into one by viewing typing derivations as outputs of sorting derivations.
  This allows us to cleanly state and prove the conservativity of our extension.
\end{abstract}

\section{Introduction}

Proof mechanization provides strong trust guarantees towards the validity of theorems.
Contrary to informal proofs, expressing a theorem and its proof formally requires absolute precision.
The resulting statement can thus become riddled with technicalities, which obscures their relation to their informal counterparts.
This work combines two approaches to type systems that significantly reduce the added complexity from formalization, namely refinement types and \ac{HOAS}.

Datasort refinement types \citep{FreemanP1991, FreemanPhD} provide ways to define subtypes (called \textit{datasorts} or just \textit{sorts}) by imposing constraints on the constructors of (inductive) types.
Intuitively, a sort $S$ refines a types $A$ if it is defined by a subset of its constructors.
The idea originated in the simply-typed setting, where refinements enhance the expressive power of the type system.
Later, Lovas and Pfenning \citep{LovasP2010, LovasPhD} extended datasort refinements to the dependently-typed Edinburgh logical framework \textsc{LF} \citep{HarperHP1993}.
They provide an equivalence between their system of refinements, \textsc{LFR}, and another extension of \textsc{LF} with proof-irrelevance.
An immediate conclusion here is that refinements do not increase the expressive power of dependently-typed calculi.
Rather, Lovas observes that refinements may significantly reduce the verbosity of mechanized proofs, which is demonstrated through several case studies \citep{LovasPhD}.

\textsc{Beluga} is a two-level programming language based on \ac{CMTT} \citep{NanevskiPP2008}.
It uses the Edinburgh logical framework \textsc{LF} \citep{HarperHP1993} as a specification logic (data-level), with an intuitionistic first-order reasoning logic (computation-level).
The data-level is embedded in the computation-level via a (contextual) box modality similar to the one in the modal logic S4.
From a logical point of view, the formula $\square A$ (read box $A$) expresses that $A$ is true under no assumptions, i.e. in the empty context.
The contextual box modality generalizes this idea to arbitrary contexts, yielding formulas of the form $[\Psi \vdash A]$ expressing that $A$ holds in context $\Psi$.
This allows us to represent \textsc{LF} objects (and types) together with a context in which they are meaningful.
To handle this representation, \textsc{LF} contexts are restricted using a notion of \textit{schema} that acts as classifiers of contexts, similarly to how types classify terms.
In addition, \textsc{LF} substitutions are first-class objects of \textsc{Beluga} and they can be used to move objects from one context to another while preserving their meaningfulness.
These features allow the expression of an \ac{OL} using \ac{HOAS} \citep{PfenningE1988} and provide several substitution lemmas for free in our mechanizations.

We present \textsc{Beluga} and its extension with refinement types in Sections \ref{sec::data} and \ref{sec::comp}, which discuss the data-level and computation-level, respectively.
The core of the extension consists of replacing the \textsc{LF} layer of \textsc{Beluga} with a variation on the \textsc{LFR} system of Lovas and Pfenning \citep{LovasP2010, LovasPhD}.
We then lift the refinement relations to the computation-level through straightforward congruence rules and show that the extension is conservative,
  meaning that every well-sorted program of our extension is well-typed in conventional \textsc{Beluga}.  
However, this result only applies if we consider \textsc{Beluga} as a general-purpose language rather than a proof environment.
This is because a \textsc{Beluga} proof is a recursive function that terminates on every input and refinements allow specifying more precise domains.
Thus, the types that we obtain from conservativity can extend the domain of a function, leading to undefined behaviour on certain inputs.
In this sense, the extension permits interpreting some partial recursive functions as proofs.
While termination is an important part of our work, we focus here on defining the refinements and leave termination checking for future work.
More details on this as well as more examples and a full definition of our system will be available in an upcoming technical report \citep{GaulinMSc}.
A version of the present paper featuring an appendix containing the fully detailed definitions of our judgments is available on the \href{https://cs.mcgill.ca/~agauli1/}{the first author's website}.


\section{Motivation}
\label{sec::motivation}

Felty et al. \citep{FeltyMP2015a} observed that binders can have various structures and that particular results rely only on particular aspects of those structures.
This leads to challenges when invoking a lemma in the proof of a theorem since the lemma may rely on simpler binding structures than the theorem.
For instance, a lemma using untyped term variables can still be useful for a theorem that uses typed term variables.
They propose a series of benchmark challenges along with solutions using multiple contexts and relations between them.
We design a new solution based on refinements for one of these benchmarks, namely the equivalence of algorithmic and declarative equalities for the untyped $\lambda$-calculus.

The first step of the mechanization is to encode the untyped $\lambda$-calculus as an \textsc{LFR} datatype:

\begin{lstlisting}[language=Beluga]
  LFR tm : type =
    | lam : (tm -> tm) -> tm
    | app : tm -> tm -> tm;
\end{lstlisting}

This syntax declares a new type called {\ttfamily\color{NavyBlue} tm} whose objects are built from the two given constructors, {\ttfamily\color{BrickRed} lam} and {\ttfamily\color{BrickRed} app}.
The constructors encode function abstraction and function application, respectively.
Next, we want to encode the judgments for declarative and algorithmic equalities:

\begin{lstlisting}[language=Beluga]
  LFR deq : tm -> tm -> type =
    | e-lam : ({x : tm} deq x x -> deq (M x) (N x)) -> deq (lam M) (lam N)
    | e-app : deq M1 N1 -> deq M2 N2 -> deq (app M1 M2) (app N1 N2)
    | e-refl : {M : tm} deq M M
    | e-sym : deq M N -> deq N M
    | e-trans : deq M1 M2 -> deq M2 M3 -> deq M1 M3;
\end{lstlisting}

Here, we exploit the dependent types of LFR to express declarative equality as a binary predicate on objects of type {\ttfamily\color{NavyBlue} tm}.
The constructors {\ttfamily\color{BrickRed} e-refl}, {\ttfamily\color{BrickRed} e-sym}, and {\ttfamily\color{BrickRed} e-trans} encode the axioms of an equivalence relation
  (reflexivity, symmetry, and transitivity, respectively),
  while the constructors {\ttfamily\color{BrickRed} e-lam} and {\ttfamily\color{BrickRed} e-app} correspond to congruence rules.
Algorithmic equality is just declarative equality without the three equivalence axioms.
As such, we define it as a refinement of {\ttfamily\color{NavyBlue} deq} rather than as a separate atomic type:

\begin{lstlisting}[language=Beluga]
  LFR aeq $\sqsubset$ deq : tm -> tm -> sort =
    | e-lam : ({x : tm} aeq x x -> aeq (M x) (N x)) -> aeq (lam M) (lam N)
    | e-app : aeq M1 N1 -> aeq M2 N2 -> aeq (app M1 M2) (app N1 N2);
\end{lstlisting}

To declare a new (atomic) sort, users must specify three things: the type which is refined, a refinement kind, and a list of constructors together with their sort.
The type must have been previously declared, the sort's kind must refine the type's kind, and each of the constructors' sort must refine their assigned type.
By using a sort instead of a type, we get to reuse the same constructors for both judgments.
This guarantees that any proof of algorithmic equality can be interpreted as a proof of declarative equality.
Thus, we get the soundness of algorithmic equality for free.

The last step in encoding the language is to define its contexts via context schemas.
The goal of a schema is to characterize the structure of contexts, which consist of tuples of assumptions rather than being flat lists.
A particular context can contain various forms of assumptions (untyped term variables, typed term variables, type variables, etc.) depending on the features of the \ac{OL} that we are mechanizing.
We call these forms of assumptions \textit{worlds} (or \textit{schema elements}).
Intuitively, worlds are to schemas what constructors are to atomic types: they specify how to construct a context of a given schema.
Our notation for schema declaration emphasizes this idea:
\begin{lstlisting}[language=Beluga]
  schema xdG =
  | xeW : block (x : tm, e_x : deq x x);
\end{lstlisting}

Now, if we have a context $\Psi$ of schema \lstinline[language=Beluga]{xdG}, then we can extend it with an additional block variable $b$ with world \lstinline[language=Beluga]{xeW},
  yielding the context $\Psi, b{:}$\lstinline[language=Beluga]{xeW}.
A refinement of schema is then obtained by selecting a subset of the worlds and refining them to sorts.
Here, we want to refine the \lstinline[language=Beluga]{deq} assumption to \lstinline[language=Beluga]{aeq}, which we do as follows :
\begin{lstlisting}[language=Beluga]
  schema xaG $\sqsubset$ xdG =
  | xeW : block (x : tm, e_x : aeq x x);
\end{lstlisting}

One advantage of this approach is that the sort of a block variable \lstinline[language=Beluga]{b:xeW} is fully hidden in the schema of the context in which it appears.
This means that a given context of schema \lstinline[language=Beluga]{xaG} can also be seen as having schema \lstinline[language=Beluga]{xdG}.
Moreover, this idea generalizes to arbitrary schemas and can go in both directions when all the worlds of the type-level schema also appear in the refinement schema.
In this case, given schemas $H \sqsubset G$ and a context $\Psi : H$, we write $\Psi^\top$ to indicate that we wish to interpret $\Psi$ as a context of schema $G$.

In the conventional solution \citep{FeltyMP2015b}, a relation between contexts of the two schemas needs to be maintained explicitly.
Here, we can simply use the refinement relation and our special $\Psi^\top$ context instead.
Let us now look at a few cases of the proof of completeness of algorithmic equality:

\begin{lstlisting}[language=Beluga]
  rec aeq-sym : ($\Psi$ : xaG) [$\Psi$ |- aeq M N] -> [$\Psi$ |- aeq N M] = ...;
  rec ceq : ($\Psi$ : xaG) [$\Psi^\top$ |- deq M N] -> [$\Psi$ |- aeq M N] =
  fn d => case d of
    | [$\Psi$ |- #b.2] => [$\Psi$ |- #b.2]
    | [$\Psi$ |- e-sym D] =>
           let [$\Psi$ |- D'] = ceq [$\Psi^\top$ |- D] in
           aeq-sym [$\Psi$ |- D']
    | [$\Psi$ |- e-lam (\x.\e. D)] =>
           let [$\Psi$, b:xeW |- E] = ceq [$\Psi$, b:xeW |- D[.., b.1, b.2]] in
           [$\Psi$ |- e-lam (\x.\e. E[..,<x;e>])]
    | ...
\end{lstlisting}
where parentheses around the context variable {\ttfamily$\Psi$ :$\ ${\color{OliveGreen}xaG}} indicate implicit quantification.

The first case is for variables, represented as the second projection on one of the block \texttt{b} in $\Psi$.
The symbol {\ttfamily\#} is merely a syntactic device to identify \texttt{b} as a variable.
This case acts in essence just like an identity function, except that the sort of the output does not match that of the input.
When we pattern match, we know that \lstinline[language=Beluga]{d} has the context $\Psi^\top$ from the sort of \lstinline[language=Beluga]{ceq}.
Since $\Psi^\top$ has schema {\ttfamily\color{Blue}xdG}, we know the world of \texttt{b} and therefore that \texttt{b.2} has sort {\ttfamily{\color{NavyBlue}deq} b.1 b.1}.
On the other hand, when we produce the output \lstinline[language=Beluga]{[$\Psi$ |- #b.2]}, then the sort of \lstinline[language=Beluga]{ceq} tells us to interpret $\Psi$ as a \lstinline[language=Beluga]{xaG},
  so we assign it the sort {\ttfamily{\color{Green}aeq} b.1 b.1}, as desired.

Next, we have the case of symmetry, which is solved by a recursive call on the subderivation, followed by a call to the relevant lemma \texttt{aeq-sym}.
We know from the sort of {\ttfamily ceq} that the recursive calls produce objects of sort $[\Psi \vdash {\color{NavyBlue}\texttt{aeq}} \ M \ N]$ with $\Psi : {\color{Blue}\texttt{xaG}}$,
  which is precisely what the lemma expects.
  
Finally, the case for $\lambda$-abstraction requires extending the context with an additional block of assumptions.
Here, it is evident that the two contexts involved are the same, except that they are interpreted in different ways.

Our solution is simple and closely resembles an informal proof of completeness of algorithmic equality.
In contrast, the conventional \textsc{Beluga} solution\footnote{Available at \url{https://github.com/pientka/ORBI}}
  requires a total of 13 additional arguments, including 7 explicit ones that must be manipulated in every case of the proof.

\section{Data-level}
\label{sec::data}

The main objective of \textsc{Beluga} is to facilitate reasoning about the properties of \acp{OL}.
To achieve this, an \ac{OL} is specified using a variant of the Edinburgh logical framework \textsc{LF} \citep{HarperHP1993}, called Contextual \textsc{LF},
  in which \textsc{LF} objects and types are always represented together with a context in which they are meaningful, that is containing all of its free variables.
The variables of an \ac{OL} are represented as \textsc{LF} variables, which allows reusing \textsc{LF}'s substitution calculus to represent substitution in the \ac{OL}.
This kind of representation is known as \ac{HOAS} and eliminates the need to prove several substitution properties.
Contextual \textsc{LFR} applies the same idea to the \textsc{LFR} system of Lovas and Pfenning \citep{LovasP2010, LovasPhD}.

We will start by reviewing Lovas and Pfenning's \textsc{LFR} \citep{LovasP2010} and discuss the numerous changes that we make to their presentation, and then we will see how refinements carry on to Contextual \textsc{LFR}.
Unfortunately, the contextual aspect cannot be cleanly separated from \textsc{LFR} since the syntax and judgments of \textsc{LFR} have to be altered during the extension.
In particular, all the judgments depend on an additional context, called the \textit{meta-context} and denoted $\Omega$ at the refinement level and $\Delta$ at the type level.
These consist of \textit{meta-variables} which may occur within \textsc{LFR} objects.
So, we maintain these aspects in our presentation of \textsc{LFR}, but defer their explanation to the last part of this section.

We follow a canonical form presentation \citep{WatkinsCPW2002} for the data-level.
This means that only normal terms are allowed, which requires the use of hereditary substitutions.
Simply put, hereditary substitutions are like ordinary substitutions except that they apply any $\beta$-reduction that appears during the process.
For instance, the substitution $[(\lambda y.y)/x](x \ 0)$ produces $0$ instead of $(\lambda y.y) \ 0$.

\subsection{\textsc{LFR}}

As previously mentioned, \textsc{LFR} extends \textsc{LF} with datasort refinement types.
The objects of \textsc{LFR} are exactly the same as in \textsc{LF} and their classifiers are separated in two levels, types $A$ and sorts $S$, which are related by a refinement relation $S \sqsubset A$.
Due to their dependencies on types, the other syntactic categories are similarly duplicated into a type-level and a sort-level related by a refinement relation.

To facilitate the extension to Contextual \textsc{LFR}, it is crucial that we apply this principle to contexts instead of using a single context containing both typing and sorting assumptions like Lovas and Pfenning \citep{LovasP2010}.
A contextual type $(\Gamma.A)$ encodes the \textsc{LF} judgment that $A$ is a well-formed type in (typing) context $\Gamma$
  and a contextual sort $(\Psi.S)$ similarly encodes the \textsc{LFR} judgment that $S$ is a well-formed sort in (sorting) context $\Psi$.
The only sensible way to establish that $(\Psi.S) \sqsubset (\Gamma.A)$ is to show that $\Psi \sqsubset \Gamma$ and that $S \sqsubset A$.
This new refinement relation can then be thought of as a relation between a sort-level judgment and a type-level judgment.

\subsubsection{Types and sorts}
%

We start by presenting the types and sorts of \textsc{LFR} and discussing the relations between them.
Both types and sorts are allowed to depend on (normal) terms $M$.
They are given by the following syntax:

\begin{center}
  \begin{tabular}{c|l|l}
    & \multicolumn{1}{c|}{Type level}  & \multicolumn{1}{c}{Refinement level}                      \\\hline
    Atomic families     & $P ::= \textbf{a} \mid P \ M$    & $Q ::= \textbf{s} \mid Q \ M \mid P$  \\
    Canonical families  & $A ::= P \mid \Pi x{:}A_1.A_2$    & $S ::= Q \mid \Pi x{:}S_1.S_2$        \\
  \end{tabular}
\end{center}

The refinement relation ultimately boils down to what the user specifies.
An atomic type family \textbf{a} is defined by its constructors and their types.
An atomic sort family $\textbf{s} \sqsubset \textbf{a}$ is then defined by selecting a subset of the constructors of \textbf{a} and assigning them sorts that refine their previously specified types.
In this sense, refinements offer a way to safely reuse constructors.
Finally, the relation is lifted to other types with simple congruence rules:
\begin{align*}
  & \infer{Q \ M \sqsubset P \ M}{Q \sqsubset P} &
  & \infer{\Pi x{:}S_1.S_2 \sqsubset \Pi x{:}A_1.A_2}{S_1 \sqsubset A_1 & S_2 \sqsubset A_2}
\end{align*}

Due to the presence of dependencies, type and sort well-formedness are non-trivial in \textsc{LFR}.
The type well-formedness judgment, $A : \texttt{type}$, coincides exactly with \textsc{LF}'s type well-formedness judgment.
It essentially just makes sure that whenever we apply an atomic family $P$ to an argument $M$, then $M$ has the type prescribed by $P$'s kind.
We could likewise consider a sort well-formedness judgment $S : \texttt{sort}$ that validates the sorts of dependencies, but instead we consider the refinement relation itself to be the sort well-formedness judgment.
This is also what was done by Lovas and Pfenning \citep{LovasP2010}, however the presence of intersection sorts in their system prevents the full separation of a sort well-formedness judgment.
This is because intersections are only allowed when both sorts refine the same type, so the refinement information must be present during the well-formedness derivation.

The refinement relation for atomic families $Q \sqsubset P$ is similar to the notion of constructor subtyping \citep{BartheF1999},
  according to which a subtyping relation $P_1 \leq P_2$ occurs between two inductive types when $P_1$ is defined by a subset of the constructors of $P_2$.
As such, it is sensible to consider a notion of subsorting (i.e. subtyping at the level of sorts) such that $Q \leq P$ whenever $Q \sqsubset P$.
In particular, a subsumption rule is admissible for refinements of atomic families.
The natural subsorting rule for function spaces would be contra-variant in the domain, while refinement of function spaces is co-variant.
As such, a subsumption principle for refinements of function spaces is not guaranteed, although it is admissible for the \textit{weak} function spaces of \textsc{LF}.

In addition to the usual typing judgment $M:A$, we have a sorting judgment, denoted $M::S$.
Sorting replicates typing similarly to how sorts replicate types syntactically.
A similar phenomenon occurs for all the other \textsc{LF} judgments (context formation, kinding, etc.).
One of our key observations is that the type-level judgments can be unified with their sort-level analogues due to their close resemblances.
For typing, this yields a judgment $M : S \sqsubset A$ that encompasses both the facts that $M:A$ and $M::S$.
Let us exemplify this by considering the rules for function applications:
\begin{align*}
  & \infer{M_1\ M_2 : [M_2/x]A_2}{M_1 : \Pi x{:}A_1.A_2 & M_2 : A_1} &
  & \infer{M_1\ M_2 :: [M_2/x]S_2}{M_1 :: \Pi x{:}S_2.S_2 & M_2 :: S_2} &
  & \infer{M_1\ M_2 : [M_2/x]S_2 \sqsubset [M_2/x]A_2}{M_1 : \Pi x{:}S_1.S_2 \sqsubset \Pi x{:}A_1.A_2 & M_2 : S_1 \sqsubset A_1} \\
  & \qquad\quad \text{(Typing)} &
  & \qquad\quad\text{(Sorting)} &
  & \qquad\qquad\qquad\;\text{(Unified)}
\end{align*}
For the remainder of our presentation, we will focus on this form of unified judgments.
In all our judgments, everything that appears to the right of the refinement symbol is considered to be an output.
For instance, in $M : S \sqsubset A$, the type $A$ is an output, which amounts to recovering a typing derivation $M:A$ from the sorting derivation $M::S$.
The actual typing rules (see Figure \ref{fig::data-typing}) are bi-directional, so we have two unified judgments, one for synthesis and one for checking.
This means that we consider neutral terms $R$ and normal terms $M$, and that we have unified judgments for synthesis ($R \Rightarrow S \sqsubset A$) and checking ($M \Leftarrow S \sqsubset A$).
We will discuss this in more details when we introduce terms in \ref{sec::data::clfr::terms}.

The other syntactic categories of \textsc{LFR} are similarly duplicated at the refinement level, except for terms which are the same at both levels since they do not contain any type information to refine.
Each category is equipped with a refinement relation that is induced by the refinement for types.
In all the judgments involving refinements, everything on the right of $\sqsubset$ can be considered as an output of the judgment.

Our presentation differs from that of Lovas and Pfenning \citep{LovasP2010} in two other ways.
First, we use an explicit embedding of types into sorts rather than an ambiguous $\top$ sort that refines every type.
This ensures that for any well-formed sort $S$, we can compute a type $A$ such that $S \sqsubset A$.
In turn, this allows us to combine the typing and sorting judgments into a single sorting judgment (see Figure \ref{fig::data-typing}).
We can then perform type-checking only when it is needed for a sorting derivation, that is when we reach a type embedded into a sort.
Moreover, our embedding is at the level of atomic families rather than canonical families.
This being said, an embedding of canonical types within canonical sorts is admissible since we can construct a $\Pi$-sort from embedded atomic type families.
This is also the case for every other syntactic category in \textsc{Beluga}.
Second, we have omitted intersection sorts $S_1 \land S_2$, which allow specifying multiple sorts for an object at once.

We note that both subsorting and intersection sorts, although useful features, can significantly slow down sort checking, much like their type-level equivalent would slow down type checking.
On the other hand, sorts themselves come at a very low cost while still offering several of the benefits of richer sort systems.
The results that we present in \ref{sec::clfr::meta} can be extended to a system supporting subsorting and intersection sorts.
Adding intersections is straightforward, but subsorting brings complication when it comes to validating coverage
  since when we pattern match on an object of sort $Q$, then we need to consider the cases coming from the constructors of $Q$ as usual, but also any additional constructor coming from a subsort of $Q$.

\subsubsection{Contexts and schemas}
%

Next, we take a closer look at \textsc{LFR} contexts and schemas.
Again, these are separated into a type-level and a refinement-level that are related by a refinement relation.
The syntax of \textsc{LFR} contexts and schemas is as follows:

\begin{center}
  \begin{tabular}{l|l|l}
                            & \multicolumn{1}{c|}{Type level}                                                & \multicolumn{1}{c}{Refinement level}                                     \\\hline
    Blocks of declarations  & $B ::= \cdot \mid \Sigma x{:}A.B$                                              & $C ::= \cdot \mid \Sigma x{:}S.C$                                        \\
    Schema elements         & $E ::= B \mid \Pi x{:}A.E$                                                     & $F ::= C \mid \Pi x{:}S. F$                                              \\
    Contexts                & $\Gamma ::= \cdot \mid \psi \mid \Gamma, x{:}A \mid \Gamma, b{:}E\cdot\vec{M}$ & $\Psi ::= \cdot \mid \psi \mid \Psi, x{:}S \mid \Psi. b{:}F\cdot\vec{M}$ \\
    Context schemas         & $G ::= \cdot \mid G + E$                                                       & $H ::= \cdot \mid H + F$
  \end{tabular}
\end{center}

Blocks of declarations represent tuples of labelled assumptions, i.e. variables.
The empty block $\cdot$ is not valid on its own, rather it is a syntactic device that indicates the end of a block.
Empty blocks are not strictly necessary for the system to work, but facilitate the theoretical development by providing a simple base case.

A schema element is a parameterized block of declarations.
While blocks express specific instances of assumptions, schema elements encode the general requirements of a particular form of assumption.
For instance, a typing assumption (informally denoted by $x : A$) is characterized by the schema element $\Pi A:\texttt{tp}.\Sigma x{:}\texttt{tm}.\Sigma t{:}\texttt{oft }x\ A. \cdot$,
  while a particular instance of this assumption would be $\Sigma x{:}\texttt{tm}.\Sigma t{:}\texttt{oft }x\ \texttt{nat}. \cdot$.
Each schema element corresponds to an inference rule for the \ac{OL}'s context formation judgment.
A schema is defined as a sum of schema elements and similarly corresponds to the \ac{OL}'s full context formation judgment.
Note that in the external syntax used in our example, every schema element was assigned a name and referred to exclusively by that name.
Here, we directly use the element's sort to avoid the need for extra premises performing signature lookups in our inference rules.

\textsc{LFR} contexts can contain two kinds of variables.
Ordinary variables, denoted by $x$, stand for an arbitrary \textsc{LFR} object of the specified type.
Block variables, denoted by $b$, stand for tuples of assumptions satisfying the specification of a schema element.
In conventional Beluga, block variables are directly assigned with a block of declaration instead of a schema element applied to some objects.
Here, we require that these objects be specified explicitly, so that they can be recovered when pattern matching on a context.
This simplifies the schema checking rules (see Appendix for a definition) since they no longer rely on unification to establish that a block extension fits a schema element.
We also allow a single context variable $\psi$ to appear on the left-most position of \textsc{LFR} contexts $\Psi$ (or $\Gamma$), but we do not consider $\psi$ as a variable of $\Psi$.
Instead, $\psi$ is a placeholder for an actual context to be substituted at a later time, so it is stored in the meta-context $\Omega$.

\begin{figure}
  $\boxed{\Omega \vdash \Psi \sqsubset \Gamma}$ -- Refinement relation for contexts
  \begin{align*}
    & \infer{\Omega \vdash \cdot \sqsubset \cdot}{\vdash \Omega \sqsubset \Gamma} &
    & \infer{\Omega \vdash \psi \sqsubset \psi}{(\psi : H) \in \Omega}
    & \infer{\Omega \vdash (\Psi, x{:}S) \sqsubset (\Gamma, x{:}A)}{\Omega \vdash \Psi \sqsubset \Gamma & \Omega; \Psi \vdash S \sqsubset A} &
    & \infer{\Omega \vdash (\Psi, b{:}F[\vec{M}]) \sqsubset (\Gamma, b{:}E[\vec{M}])}{\Omega \vdash \Psi \sqsubset \Gamma & \Omega; \Psi \vdash F \sqsubset E}
  \end{align*}
  \caption{Refinement relations for contexts and schemas}
  \label{fig::data-refinement}
\end{figure}

The refinement relations for blocks, schema elements, and schemas are very simple (see Appendix).
For schema elements, we just check one sort at a time, starting with the parameters and then the assumptions in the block.
Refinements of contexts are similarly checked one assumption at a time (see Figure \ref{fig::data-refinement}).
For block assumptions, we require the same parameters to be used to instantiate the schema elements on both sides of the refinement relation.
The relation on schemas is similarly simple, but we take care not to allow duplicate schema elements in $G$ (or in $H$ for that matter).
We do this mainly because duplicate elements serve no purpose in practice, but also to highlight the fact that multiple elements of $H$ can refine the same element of $G$.

Contexts are validated using the schema checking judgments $\Omega \vdash \Psi : H \sqsubset G$ (at the sort-level) and $\Delta \vdash \Gamma : G$ (at the type-level).
Assigning a schema $H$ to a context $\Psi$ requires that all the assumptions in $\Psi$ match one of the schema elements in $H$ (see Appendix).
This means that all the assumptions in $\Psi$ are of the form $b{:}F[\vec{M}]$, hence there is no rule associated to single variables $x{:}A$.
The empty context checks against any well-formed schema.
For context extensions, we use an auxiliary judgment $\Omega \vdash \vec{M} : F > D$ that checks the terms in $\vec{M}$ against the parameters of $F$ one at a time.
In the end, it produces the block of declarations $D$ obtained by $\beta$-reducing $F[\vec{M}]$.
Recall that $A$ is an output of the judgment $\Omega; \cdot \vdash M \Leftarrow S \sqsubset A$ and that $C$ is an output of $\Omega \vdash D \sqsubset C$,
  so we do not need to know them in advance in order to validate the premises.

\subsubsection{Terms and substitutions}
\label{sec::data::clfr::terms}

Now that we have discussed the classifiers of contextual \textsc{LFR}, let us look at the objects that they classify.
We distinguish two kinds of objects, namely terms and substitutions.
Terms are classified by sorts (and types), while substitutions are classified by contexts.
Only normal forms are allowed at the data-level, and this is enforced with a canonical form presentation \citep{WatkinsCPW2002}.
The syntax is as follows:

\begin{center}
  \begin{tabular}{lrllrl}
    Neutral term  &\quad $R ::=$       & $\textbf{c} \mid x \mid b.k \mid R \ M$      &\qquad\qquad
    $n$-ary tuple &\quad $\vec{M} ::=$ & $\cdot \mid M;\vec{M}$                       \\
    Normal term   &\quad $M ::=$       & $R \mid u[\sigma] \mid \lambda x.M$          &\qquad\qquad
    Substitution  &\quad $\sigma ::=$  & $\cdot \mid \texttt{id}_\psi \mid \sigma, M \mid \sigma, \vec{M}$
  \end{tabular}
\end{center}

The separation of terms into neutral and normal ensures that no $\beta$-reduction can be done by preventing applications of $\lambda$-abstractions.
The typing rules (see Figure \ref{fig::data-typing}) will also guarantee that all terms are $\eta$-long.
$n$-ary tuples of normal terms are crucially used in substitutions to replace block variables $b$.
Since $b$ is always used in a projection $b.k$, the substitution $[\vec{M}/b]$ needs to extract the $k^{th}$ projection of the $n$-ary tuple
  in order to avoid expressions of the form $\vec{M}.k$, which are undefined by our grammar (and not normal).
This coincides nicely with the idea of hereditary substitution and can be added with only minor modifications to their definition.
Substitutions can also contain individual terms, which are used to replace individual variables $x$.
Finally, $\texttt{id}_\psi$ is the identity substitution for the context variable $\psi$.
Note that substitutions have the same structure as their domain, hence it is not specified explicitly.

\begin{figure}[!ht]
  $\boxed{\Omega; \Psi \vdash M \Leftarrow S \sqsubset A}$ and $\boxed{\Omega; \Psi \vdash R \Rightarrow S \sqsubset A}$ -- Bi-directional typing
  \begin{align*}
    & \infer{\Omega; \Psi \vdash u[\sigma] \Leftarrow [\sigma]Q \sqsubset [\sigma]P}
            {
              (u : \Psi'.P) \in \Omega
              &
              \Omega; \Psi \vdash \sigma : \Psi' \sqsubset \Gamma
            }
  \end{align*}
  \vspace{-24pt}
  \begin{align*}
    & \infer{\Omega; \Psi \vdash b.k \Rightarrow S \sqsubset A}
            {
              (b : F[\vec{M}]) \in \Psi
              &
              \Omega \vdash \vec{M} : F > C
              &
              \Omega; \Psi \vdash b : C \gg_1^k S
              &
              \Omega; \Psi \vdash S \sqsubset A
            }
  \end{align*}

  $\boxed{\Omega; \Psi_1 \vdash \sigma : \Psi_2 \sqsubset \Gamma_2}$ -- $\sigma$ is a well-formed substitution from $\Psi_2$ to $\Psi_1$
  \begin{align*}
    \infer{\Omega; \Psi_1 \vdash (\sigma, \vec{M}_1) : (\Psi_2, b{:}F[\vec{M}_2]) \sqsubset (\Gamma_2, b{:}E[\vec{M}_2])}
          {
            \Omega; \Psi_1 \vdash \sigma : \Psi_2 \sqsubset \Gamma_2
            &
            \Omega; \Psi_2 \vdash \vec{M}_2 : F > D
            &
            \Omega; \cdot \vdash F \sqsubset E
            &
            \Omega; \Psi_1 \vdash \vec{M}_1 \Leftarrow D
          }
  \end{align*}
  \caption{Bi-directional typing rules}
  \label{fig::data-typing}
\end{figure}

Neutral terms consist of constants \textbf{c}, single \textsc{LFR} variables $x$, projections of \textsc{LFR} block variables $b.k$, and function applications of neutral terms to normal terms $R\ M$.
The sort synthesis rules for constants, single variables, and function applications are standard and coincide with those of Lovas and Pfenning \citep{LovasP2010}.
Blocks of variables $b$ are not valid \textsc{LFR} objects on their own, instead they are always used in projections.
To synthesize the sort of a projection $b.k$, we first retrieve its classifying world $F[\vec{M}]$ from the context,
  then we compute the block $D$ that it corresponds to via the judgment $\Omega \vdash \vec{M} : F > D$,
  and finally we extract the $k^{th}$ component of $D$ using the auxiliary judgment $\Omega; \Psi \vdash b : C \gg_i^k$.
Normal terms are either neutral terms or $\lambda$-abstraction, and the sort-checking rules are standard.
  
Well-formedness of substitutions is validated with the judgment $\Omega; \Psi_1 \vdash \sigma : \Psi_2 \sqsubset \Gamma$, where $\Psi_2$ is the domain and $\Psi_1$ the range of $\sigma$.
The empty substitution has domain $\cdot$ and range any context $\Psi_1$, so it allows weakening a closed object to any context.
Substitution extension with a single term ($\sigma, M$) are validated against context extensions with a single variable ($\Psi_2, x{:}S$) in the usual way.
The substitution $\sigma, \vec{M}$ is used to substitute the $n$-ary tuple $\vec{M}$ for a block variable $b$.

\subsubsection{Meta-theory}
\label{sec::clfr::meta}

Our main result concerning \textsc{LFR} is that for any refinement-level derivation, there is a corresponding type-level derivation.
In particular, well-sorted terms are also well-typed, which means that our extension does not provide any new meaningful terms, i.e. that it is conservative.
The statement of the theorem relies on type-level \textsc{LFR} judgments that we have not yet discussed due to their similarities with analogous refinement-level judgments and to the fact we did not change them.
Let us then first recapitulate the important refinement-level judgments and mention their type-level analogues:

\begin{center}
  \begin{tabular}{l|l|l}
    \multicolumn{1}{c|}{Judgment} & \multicolumn{1}{c|}{Type-level}      & \multicolumn{1}{c}{Refinement-level}       \\\hline
    Type formation                & $\Delta; \Gamma \vdash A \Leftarrow \texttt{type}$ & $\Omega; \Psi \vdash S \sqsubset A$        \\
    Type checking                 & $\Delta; \Gamma \vdash M \Leftarrow A$     & $\Omega; \Psi \vdash M \Leftarrow S \sqsubset A$ \\
    Type synthesis                & $\Delta; \Gamma \vdash R \Rightarrow A$     & $\Omega; \Psi \vdash R \Rightarrow S \sqsubset A$ \\
    Schema checking               & $\Delta \vdash \Gamma : G$           & $\Omega \vdash \Psi : H \sqsubset G$
  \end{tabular}
\end{center}

The same separation occurs for any other judgment of the system.
In particular, every refinement relation that we have introduced corresponds to the type-level formation judgment of the associated syntactic category, like for types.
The rules defining a type-level judgment are roughly the same as those defining its refinement-level analogue, except that the refinement-level information is replaced by type-level information.
Now, we can formulate the conservativity theorem for \textsc{LFR} as follows:

\begin{theorem}[Conservativity for data-level] \hfill
  \begin{enumerate}
  \item If $\Omega; \Psi \vdash S \sqsubset A$, then there are $\Delta$ and $\Gamma$ such that:
    \vspace{-12pt}
    \begin{multicols}{3}
      \begin{enumerate}
      \item $\vdash \Omega \sqsubset \Delta$,
      \item $\Omega \vdash \Psi \sqsubset \Gamma$,
      \item $\Delta; \Gamma \vdash A \Leftarrow \texttt{type}$.
      \end{enumerate}
    \end{multicols}
    \vspace{-12pt}
  \item If $\Omega; \Psi \vdash M \Leftarrow S \sqsubset A$, then there are $\Delta$ and $\Gamma$ such that:
    \vspace{-12pt}
    \begin{multicols}{3}
      \begin{enumerate}
      \item $\vdash \Omega \sqsubset \Delta$,
      \item $\Omega \vdash \Psi \sqsubset \Gamma$,
      \item $\Delta; \Gamma \vdash M \Leftarrow A$.
      \end{enumerate}
    \end{multicols}
    \vspace{-12pt}
  \item If $\Omega \vdash \Psi: H \sqsubset G$, then there are $\Delta$ and $\Psi$ such that:
    \vspace{-12pt}
    \begin{multicols}{3}
      \begin{enumerate}
      \item $\vdash \Omega \sqsubset \Delta$,
      \item $\Omega \vdash \Psi \sqsubset \Gamma$,
      \item $\Delta \vdash \Gamma : G$.
      \end{enumerate}
    \end{multicols}
  \end{enumerate}
\end{theorem}

The proof is discussed in \ref{sec::data::clfr::meta}.
For now, we simply observe that the close resemblance between type-level and refinement-level judgments, combined with the fact that we can extract type-level derivations from refinement-level ones,
  suggests that we can lift the refinement relations to the level of \textsc{LFR} judgments.
This idea will be reinforced by the refinement relations on contextual types.

\subsection{Contextual \textsc{LFR}}

The contextual layer (also known as the meta-layer \citep{CaveP2012}) unifies the different kinds of objects and classifiers of the data-level into unique constructs.
This facilitates function abstraction at the computation-level since otherwise each kind of object would need a special kind of function space.
As before, our classifiers are separated into types and refinement types.
Moreover, since contextual objects include \textsc{LFR} contexts, we naturally obtain a refinement relation for objects as well.
The syntax is as follows:

\begin{center}
  \begin{tabular}{c|l|l}
                       & \multicolumn{1}{c|}{Type level}                               & \multicolumn{1}{c}{Refinement level}                   \\\hline
    Contextual types   & $\mathcal{A} ::= \Gamma.P \mid \Gamma.\Gamma' \mid G$                  & $\mathcal{S} ::= \Psi.Q \mid \Psi.\Psi' \mid H$                 \\
    Contextual objects & $\mathcal{M} ::= \hat{\Gamma}.R \mid \hat{\Gamma}.\sigma \mid \Gamma$  & $\mathcal{N} ::= \hat{\Psi}.R \mid \hat{\Psi}.\sigma \mid \Psi$ \\
    Meta-contexts      & $\Delta ::= \cdot \mid \Delta, X{:}\mathcal{A}$                        & $\Omega ::= \cdot \mid \Omega, X{:}\mathcal{S}$                 \\
    Meta-substitutions & $\rho ::= \cdot \mid \rho, \mathcal{M}$                                & $\theta ::= \cdot \mid \theta, \mathcal{N}$                     \\
    Meta-variables     & \multicolumn{2}{c}{$X ::= u \mid \psi$}
  \end{tabular}
\end{center}

Contexts with hats ($\hat{\Gamma}, \hat{\Psi}$) are called \textit{erased} and contain no type or sort information, so they consist only of variables.
Erased contexts are sufficient in this setting since the LF neutral objects and LF substitution do not refer to any type or sort information present in the context.
Note that if $\Psi \sqsubset \Gamma$, then $\hat{\Psi} = \hat{\Gamma}$.
This means that if $\mathcal{N} \sqsubset \mathcal{M}$ are not just contexts, then $\mathcal{N} = \mathcal{M}$.
Accordingly, refinements of meta-objects only provides information when contexts are used as objects.

Note that we only allow atomic \textsc{LFR} sorts to occur in the contextual sort $\Psi.Q$ (and similarly at the type-level).
This is not a real limitation of the system since the sort $\Psi.\Pi x{:}S_1.S_2$ would be isomorphic to $(\Psi,x{:}S_1).S_2$.
Similarly, we allow only neutral terms in contextual objects $\hat{\Psi}.R$, but this does not impact the expressiveness.

A meta-context can contain two kinds of variables, each associated with one of the three possible forms of contextual types.
The meta-variable $u$ stands for an \textsc{LFR} term, so it is given the contextual sort $\Psi.Q$.
Meta-variables must be associated with an \textsc{LFR} substitution $\sigma$ before being used in \textsc{LFR} terms as $u[\sigma]$.
In this setting, $\sigma$ is delayed until a meta-substitution is applied to replace $u$.
The other form of meta-variables are context variables $\psi$, which are assigned a context schema $H$.
Context variables are used to quantify over contexts at the computation-level.
The system can also be extended with support for substitution variables \citep{Pientka2008, CaveP2013}.

A meta-substitution is similar to an ordinary substitution, except that it substitutes contextual objects for contextual variables.
We denote the application of a meta-substitution using double square brackets.
For instance, $\llbracket \theta \rrbracket M$ applies the meta-substitution $\theta$ to the \textsc{LFR} normal term $M$.
A full definition of this operation was given by Cave and Pientka \citep{CaveP2013}.

The refinement relation for contextual types is obtained by lifting the corresponding refinement relations (developed previously).
Similarly, the refinement relation for meta-objects is obtained by lifting the refinement relation on \textsc{LFR} contexts.
For example, the natural refinement rule for $\Psi.Q$ is the following:
\[
\infer{\Omega \vdash \Psi.Q \sqsubset \Gamma.P}
      {
        \Omega \vdash \Psi \sqsubset \Gamma
        &
        \Omega; \Psi \vdash Q \sqsubset P
      }
\]
In a sense, this raises the refinement relation to the level of \textsc{LFR} judgments.
It is helpful to pursue this idea further by formulating our rules for the meta- and computation-level as a refinement relation between judgments.
The above rule would then become:
\[
\infer{(\Omega \vdash \Psi.Q) \sqsubset (\Delta \vdash \Gamma.P)}
      {
        \vdash \Omega \sqsubset \Delta
        &
        \Omega \vdash \Psi \sqsubset \Gamma
        &
        \Omega; \Psi \vdash Q \sqsubset P
      }
\]
This judgment $(\Omega \vdash \mathcal{S}) \sqsubset (\Delta \vdash \mathcal{A})$ can then serve as both a type well-formedness and a refinement (i.e. sort well-formedness) judgment.
We can similarly unify the sorting and typing judgments, the context well-formedness and refinement judgments, and so on (see Appendix).
In all of these judgments, the type-level part (everything on the right of $\sqsubset$) can be taken independently for the rest, in which case it defines the usual judgment of conventional \textsc{Beluga}.
On the other hand, we can also consider the type-level part to be an output, which highlights the fact that type-level judgments do not need to be validated prior to their sort-level analogues.

\subsubsection{Meta-theory}
\label{sec::data::clfr::meta}

The new refinement relation between judgments also facilitates formulating our conservativity result since we have all the type information available by assumption:

\begin{theorem}[Conservativity for contextual layer] \hfill
  \label{thm::meta-conservativity}
  \begin{enumerate}
  \item If $(\Omega \vdash \mathcal{S}) \sqsubset (\Delta \vdash \mathcal{A})$, then $\Delta \vdash \mathcal{A}$.
  \item If $(\Omega \vdash \mathcal{N} : \mathcal{S}) \sqsubset (\Delta \vdash \mathcal{M} : \mathcal{A})$, then $\Delta \vdash \mathcal{M} : \mathcal{A}$.
  \item If $(\Omega_1 \vdash \theta : \Omega_2) \sqsubset (\Delta_1 \vdash \rho : \Delta_2)$, then $\Delta_1 \vdash \rho : \Delta_2$.
  \end{enumerate}
\end{theorem}

Conservativity for the contextual layer is proven simultaneously with conservativity for \textsc{LFR} due to inter-dependencies between the two.
The proof is a straightforward induction on the given derivation.

An important advantage of our formulation of refinements as a relation on judgments is that it eliminates the need for several lemmas, in particular substitution properties.
We note that to obtain the full benefits of the approach, we must also formulate the refinements for \textsc{LF} in this style, as otherwise the lemmas are still needed for conservativity of \textsc{LFR}.

\section{Computation-level}
\label{sec::comp}

\textsc{Beluga}'s computation-level is an \textsc{ML}-style functional programming language with support for pattern matching over contextual objects.
It features an indexed function space, so that types are allowed to depend only on data-level objects.
Contextual objects and types are embedded in the computation-level via a box modality.

\subsection{Computation-level refinements}

In our extension, the computation-level is separated into a type layer and a refinement layer, just like the data-level.
Since contextual objects can occur in computation-level expression, we maintain a refinement relation for expressions in addition to all other syntactic categories.
Our presentation is inspired by the one of Pientka and Abel \citep{PientkaA2015}, but differs in two important ways.
First, we do not consider recursion since it complicates the syntax of patterns significantly.
Specifically, valid recursive calls have to be specified as part of every pattern (although they can be inferred, so users do not need to provide them explicitly).
Without recursion, patterns are just (boxed) contextual objects.
Second, our sorting (and typing) rules do not require coverage for
pattern matching. The syntax of the computation-level is the following:

\begin{center}
  \begin{tabular}{l|c|rl}
                & Type level & \multicolumn{2}{c}{Refinement level}                                                                   \\\hline
    Types       & $\tau$     & $\zeta ::=$ & $[\mathcal{S}] \mid \zeta_1 \rightarrow \zeta_2 \mid \Pi X{:}\mathcal{S}.\zeta$                                    \\
    Contexts    & $\Xi$      & $\Phi ::=$  & $\cdot \mid \Phi, y{:}\zeta$                                                              \\
    Expressions & $e$        & $f ::=$     & $[\mathcal{N}] \mid {\color{violet}\texttt{fn}}\ y{:}\zeta \Rightarrow e \mid e_1 \ e_2 \mid {\color{violet}\texttt{mlam}}\ X{:}\mathcal{S} \Rightarrow e \mid e \ \mathcal{N}$  \\
                &            &             & $\hphantom{m} \mid {\color{violet}\texttt{let}}\ [X] = e_1 \ {\color{violet}\texttt{in}}\ e_2 \mid {\color{violet}\texttt{case}}^\zeta\ [\mathcal{N}]\ {\color{violet}\texttt{of}}\ \vec{c}$       \\
    Branches    & $b$        & $c ::=$     & $\Omega;[\mathcal{N}] \mapsto e$
  \end{tabular}
\end{center}

Meta-types, -sorts, and -objects are embedded into the computation level via a (contextual) box modality, denote $[\mathcal{S}]$ (for sorts).
The elimination form for the modality is given by the {\color{violet}\texttt{let}} expressions: an expression $e_1 : [\mathcal{S}]$ is unboxed as the meta-variable $X$, which may then be used in the expression $e_2$.

We distinguish two kinds of function spaces, the simple function space $\zeta_1 \rightarrow \zeta_2$ and the dependent function space $\Pi X{:}\mathcal{S}.\zeta$.
So, dependencies are restricted to objects from the index domain, which provides strong reasoning power over the index domain without all the difficulties of full dependent types.

The language also supports pattern matching on meta-objects through the use of {\color{violet}\texttt{case}} expressions.
While we do not allow pattern matching on arbitrary expressions, any expression that has a box sort can be matched against by first unboxing it with a {\color{violet}\texttt{let}} expression and then matching on the new variable.
The sort superscript $\zeta$ in {\color{violet}\texttt{case}} expression corresponds to the sort invariant that must be satisfied by all the branches in $\vec{c}$.
We require that invariants have the form $\Pi\Omega_1.\Pi X_0:\mathcal{S}_0.\zeta_0$.
Intuitively, a branch $\Omega;[\mathcal{N}] \mapsto e$ satisfies the invariant $\Pi\Omega_1.\Pi X_0{:}\mathcal{S}_0.\zeta_0$
  if $\mathcal{N}$ has sort $\mathcal{S}_0$ and $e$ has sort $\llbracket \mathcal{N}/X_0 \rrbracket\zeta_0$, where $\mathcal{N}$, $e$, and their sorts can depend on $\Omega$.

The judgments for the computation-level have a similar structure as those for contextual \textsc{LFR}.
In particular, type-level and refinement-level judgments are performed simultaneously, with the type-level judgment seen as an output of the simultaneous judgment.
Since the derivations produced on both sides of the refinement relation are almost exactly the same, we give the rules with only the refinement part.
For instance, sorting and typing is expressed as $(\Omega; \Phi \vdash f : \zeta) \sqsubset (\Delta; \Xi \vdash e : \tau)$, but we define only $\Omega; \Phi \vdash f : \zeta$ for conciseness.
We focus here on the rules related to pattern matching.
The remaining rules are standard and can be found in the appendix.
The rule for {\color{violet}\texttt{case}}-expressions is the following:
\[
\infer{\Omega; \Phi \vdash ({\color{violet}\texttt{case}}^\zeta\ [\mathcal{N}]\ {\color{violet}\texttt{of}}\ \vec{c}) : \llbracket \rho, \mathcal{N}/X_0 \rrbracket \zeta_0}
      {
        \zeta = \Pi \Omega_0. \Pi X_0{:}\mathcal{S}_0.\zeta_0
        &
        \Omega \vdash \rho : \Omega_0
        &
        \Omega \vdash \mathcal{N} : \llbracket \rho \rrbracket \mathcal{S}_0
        &
        \Omega; \Phi \vdash c : \zeta\ (\text{for all } c \in \vec{c})
      }
\]
The important part of this rule is the last premise, which requires validating that every branch satisfies the given invariant.
This is achieved with the judgment $\Omega; \Phi \vdash c : \zeta$ defined by the following rule:
\[
\infer{\Omega; \Phi \vdash (\Omega_0; [\mathcal{N}_0] \mapsto f) : \Pi \Omega_1. \Pi X_0{:}\mathcal{S}_0.\zeta_0}
      {
        \Omega_0 \vdash \mathcal{N}_0 : \mathcal{S}_0
        &
        \Omega, \Omega_0 \vdash \mathcal{S} \doteq \mathcal{S}_0 / (\rho, \Omega')
        &
        \Omega'; \llbracket \rho \rrbracket \Phi \vdash \llbracket \rho \rrbracket f : \llbracket \rho \rrbracket \zeta_0
      }
\]
where the judgment $\Omega \vdash \mathcal{S} \doteq \mathcal{S}' / (\rho, \Omega')$ denotes (meta-type) unification.
The main difficulty of unification is unifying the dependencies on \textsc{LF}(\textsc{R}) terms.
Since we have not modified terms in our extension, the unification algorithm of Pientka and Pfenning \citep{PientkaP2003} still applies.

The conservativity results discussed in section \ref{sec::data::clfr::meta} carry over to the computation-level via straightforward inductions.
In particular, every sorting derivation has an analogous typing derivation.

\section{Related work}
\label{sec::related}

\subsection{Refinement types}
Various forms of refinement types have been used to solve various problems.
Our work is inspired by the datasort tradition that was initiated by Freeman and Pfenning \citep{FreemanP1991, FreemanPhD} for \textsc{MiniML}, a monomorphic fragment of \textsc{Standard ML}'s core language.
Their system uses refinement type inference so that users do not need to provide annotations, but the inferred sorts are often intersections with some undesired components.

Davies \citep{DaviesPhD}, who coined the term datasort, extended this work to the full \textsc{Standard ML} language (including modules).
They ditched sort inference in favour of a bi-directional sort-checking algorithm, so that only the desired sorts are used by the compiler.
Unfortunately, even sort-checking is untenable in the presence of intersection (at least in theory).
The compiler needs to choose the correct branch of an intersection when synthesizing sorts for neutral expressions, making sort-checking PSPACE-hard \citep{Reynolds1997}.

Jones and Ramsay \citep{JonesR2021} used refinement types to validate termination of functional programs in the presence of non-exhaustive pattern matching.
Their notion of an \textit{intensional} refinement is obtained by removing some of the constructors from a datatype, but the remaining constructors cannot be assigned new sorts.
Instead, a constructor $\textbf{c} : A$ selected for the sort $\textbf{s} \sqsubset \textbf{a}$ is assigned the sort $S$ obtained by replacing every occurrence of $\textbf{a}$ in $A$ by $\textbf{s}$.
Intensional refinements are weaker than datasorts, but they have full type and refinement inference for a polymorphic \textsc{ML}-style language with algebraic datatype.
Their main ideas would also be sufficient to encode our example from section \ref{sec::motivation}.

Another important (and perhaps more common) approach is index refinements, first introduced by Xi and Pfenning \citep{XiPhD, XiP1999} for the core language of \textsc{Standard ML}.
They design a family of dependently-typed \textsc{ML}-style languages parameterized by an arbitrary index domain $C$, called $\textsc{DML}(C)$.
Refinements are obtained by allowing quantification over the index domain, which intuitively corresponds to having a refinement relation $\Pi x{:}S.A \sqsubset A$, where $S \in C$.
In this way, most difficulties of dependent types can be avoided, similarly to how we avoid them in \textsc{Beluga}'s computation-level.
Datasort refinements and index refinements were combined by by Dunfield \citep{DunfieldPhD}, yielding an extension of \textsc{DML} with intersections.

An important development of this approach came in the form of logically qualified (or liquid) types \citep{RondonKR2008}, this time as an extension of \textsc{OCaml}.
In this methodology, a refinement is expressed as $\{x:\tau \mid P(x)\}$, where $\tau$ is a type and $P$ is a boolean-valued function over $\tau$.
The type $\tau$ can then be seen as the refinement $\{x:\tau \mid \texttt{true}\}$ and this allows combining typing and sorting into one judgment, much like we have done for datasort refinements.

To our knowledge, there is currently no work on index refinements for dependently-typed languages.
A series of papers culminated in the lax logical framework with side conditions $\textsc{LLF}_{\mathcal{P}}$ \citep{HonsellLMS2017}, which contains types similar in spirit to liquid types.
However, $\textsc{LLF}_{\mathcal{P}}$ uses a notion of \textit{lock} types that is based on monads instead of refinements.
This being said, side conditions in $\textsc{LLF}_{\mathcal{P}}$ can be interpreted as proof irrelevance, which coincides with Lovas and Pfenning's interpretation of sorts in \textsc{LFR} as proof irrelevance \citep{LovasP2010}.
$\textsc{LLF}_{\mathcal{P}}$ allows more side conditions than \textsc{LFR}, but also puts a heavier proof burden on the user.

\subsection{Proof environments}
Although \ac{HOAS} representations offer undeniable benefits, there remains few proof environments that support it.
\textsc{Twelf} \citep{PfenningS1999} uses it in its implementation of \textsc{LF}, but all contexts are implicit and this prevents representing context relations \citep{FeltyMP2015b}.
Context schemas emerged from work on \textsc{Twelf} \citep{SchurmannPhD}, although their notion is more restrictive than ours.
The judgments have to be indexed by the unique schema to which the ambient context is known to adhere and there is no way to consider two derivations with different schemas simultaneously.

\textsc{Hybrid} \citep{FeltyM2012} only partially achieves the goals of \ac{HOAS}:
  contexts and substitution are inherited from the meta-language, but substitution properties in the \ac{OL} still need to be proven by hand.
However, it has the advantage of being built in well-known logics from which it naturally inherits consistency.
  
\textsc{Abella} \citep{Gacek2008} does provide the full power of \ac{HOAS} (called $\lambda$-tree syntax in their terminology).
Variables are represented in the specification logic through the use of the $\nabla$-quantifier (pronounced nabla).
Contexts are represented as lists of $\nabla$-quantified variables and schemas as types depending on these lists.
\textsc{Adelfa} \citep{Southern2021} is inspired by \textsc{Abella}, but uses a specification language that is more closely related to \textsc{LF}.
It also improves \textsc{Abella} with a built-in notion of schemas.

\section{Conclusion}

We have developed an extension of Beluga with datasort refinement types.
While datasort refinements are mainly used to provide subtyping and intersection types to a language, refinements in the setting of Beluga offers the potential to express proofs more succinctly.
Our extension mainly focused on the notion of refinement schemas,
  which allow extracting more precise information about contexts, much like refinements extract more precise properties (than types) about objects.
In particular, refinement schemas are useful to deal with a special kind of context relations, namely those when the assumptions in two contexts are related by refinements.

\subsection{Future work}
There are a number of avenues that we plan to explore in the future.
First, refinements allow validating the correctness of functions containing non-exhaustive pattern matching thereby supporting modular proof development.
A natural next step is therefore to develop a coverage and termination checker for Beluga with refinement types.
Due to the close similarity between types and their refinements, it is reasonable to expect that we can adapt \citep{PientkaA2015}.
However, challenges emerge from the fact that objects can have multiple sorts (but only one type), especially when it comes to unification.
This will in particular impact how we check for coverage.

Second, the refinement relations that we described for schemas and schema elements are limited by the fact that our meta-theory requires sorts to refine only one type.
In particular, the relation $D \sqsubset C$ for blocks requires that $D$ and $C$ have the same length.
A more interesting relation would allow $D$ to contain more assumptions than $C$.
This is reminiscent of the subtyping principle that adding fields to a record type produces a subtype.
Another limitation is that $H \sqsubset G$ can only be established when every element of $G$ is refined by some element of $H$.
Allowing more elements to appear in $G$ would also be meaningful, especially given that every context in $G$ is also in $G + E$ for any element $E$.
With these two modifications, we could represent another class of context relations that Felty et al. \citep{FeltyMP2015a} calls \textit{linear extensions}.
Consequently, our next goal is to provide a more flexible form of refinements and to modify our proof of conservativity so that it no longer relies on type uniqueness for refinements.

\bibliographystyle{eptcs}
\bibliography{bibliography}


\end{document}